# Orientation Dynamics of Asymmetric Rotors Using Random Phase Wave Functions

Shimshon Kallush[1†] and Sharly Fleischer[2‡]

[1] *Dept. of Physics and Optical Engineering, ORT Braude College, PO Box 78, 21982 Karmiel, Israel.*
[2] *Department of Chemical Physics, Tel-Aviv University, Tel Aviv 69978, Israel.*

Abstract:

Intense terahertz-frequency pulses induce coherent rotational dynamics and orientation of polar molecular ensembles. Exact numerical methods for rotational dynamics are computationally not feasible for the vast majority of molecular rotors - the asymmetric top molecules at ambient temperatures. We exemplify the use of Random Phase Wave Functions (RPWF) by calculating the terahertz-induced rotational dynamics of sulfur dioxide ($SO_2$) at ambient temperatures and high field strengths and show that the RPWF method gains efficiency with the increase in temperature and in the THz-field strengths. The presented method provides wide-ranging computational access to rotational dynamical responses of molecules at experimental conditions which are far beyond the reach of exact numerical methods.

Laser induced coherent molecular rotation has been actively researched for more than two decades and has enabled a wide range of possibilities in ultrafast spectroscopy and control of gas phase molecular dynamics as well as various fundamental physical phenomena that manifest in molecular ensembles with anisotropic angular distributions[1,2,3]. Most of the efforts have been put to molecular *alignment* where gas phase molecules attain highly anisotropic distributions such that their inter-nuclear axes lie along a chosen lab-frame direction (typically referred to as the z-lab axis). In recent years, significant efforts have been put toward molecular *orientation* in which the permanent dipoles of the molecules point preferentially in a chosen lab-frame axis (positive or negative lab-frame z axis). In this situation the macroscopic inversion symmetry of the medium is lifted momentarily and provides the necessary condition for various nonlinear phenomena induced by even number of light-matter interactions such as even harmonic generation[4] and directional molecular ionization/dissociation[5]. The techniques used for orientation rely on mixed field ($\omega+2\omega$) rotational excitation [4,5,6], mixed DC+Near-IR fields [7, 8, 9] and intense terahertz frequency fields [10, 11, 12, 13].

The foundations of the quantum rotational dynamics of linear, symmetric-top and asymmetric-top molecules have been actively reinforced in the last two decades [1, 14, 15, 16, 17, 18, 19, 20, 21]. However, while most of the chemically- and biologically-significant molecules are asymmetric-tops, the vast majority of theoretical and experimental research in the field has focused on the first two groups. Linear and symmetric-top molecules are typically modeled as quantum mechanical rigid rotors with one moment of inertia and correspondingly, one rotational degree of freedom that exhibit periodic rotational evolution known as quantum revivals. Numerical simulations of asymmetric tops are significantly more computationally-demanding than linear and symmetric-top molecules, due to the existence of three coupled rotational degrees of freedom associated with three different principal moments of inertia. This increases the density of states and a large number of thermally occupied rotational states are required, even at low temperatures, to correctly represent the initial thermal ensemble. For small molecules with initial rotational temperatures in the few-Kelvin regime, the computational complexity of the simulation is reduced by restricting the Hilbert space to few tens of initial eigenstates that can be propagated directly at reasonable computational effort [22]. However, simulating for ambient ensemble temperatures remains a central challenge in asymmetric rotational dynamics and in large quantum systems in general.

A promising method for calculating the dynamics of large open quantum systems relies on the fact that the quantum phase relations in such large thermal ensembles are effectively random. The method, termed Random Phase Wave Functions (RPWF) have been successfully utilized in calculating linear response functions of free electrons [23], simulating dissipative phenomena via the surrogate Hamiltonian method [24], and analyzing the inelastic atom-surface scattering at finite temperatures [25]. Recently, the method was utilized to calculate the photo-association of hot magnesium atom pairs [26, 27], and to perform multi-configuration time-dependent Hartree-Fock simulations [28].

In this paper, we implement and explore the RPWF for simulating the rotational dynamics of an asymmetric top model molecule ($SO_2$) at initial ensemble temperatures that cannot be simulated otherwise. We study the accuracy and predictability power of the RPWF method by comparing to exact molecular dynamics calculation for different initial temperatures and field strengths of a single-cycle terahertz (THz) pulse that induces rotational dynamics and transient orientation of the molecular ensemble. The rotational responses of $SO_2$ by ultrashort, nonresonant near-IR pulses have been explored recently by several groups [29, 30, 31] and are revisited here with intense, single-cycles THz fields that yield molecular orientation as a test-bed for the RPWF method. We note that simulating THz-induced molecular orientation at high temperatures is of special value from the experimental point of view since the resonant THz - dipole interaction depends on the overlap between the rotational transition spectrum and the typical frequencies of the THz pulse [32]. However, the presented method is capable of computing the rotational responses of molecules at high temperatures, induced by (non-resonant) optical and/or (resonant) THz fields, which are in many cases experimentally inherent to the observed phenomena such as high harmonic generation spectroscopy, laser filamentation and control of the optical properties of an extended medium, etc.

The calculations were performed for SO$_2$ molecules, described by three rotational constants $(A = 2.028 cm^{-1}, B = 0.3442 cm^{-1}, C = 0.2935 cm^{-1})$ for the a,b,c molecular axes respectively, and a permanent dipole moment $\mu = 1.62\ Debye$ in the direction of the molecular $a$-axis (Fig. 1d). The applied field is taken as a single-cycle THz pulse with Gaussian pulse envelope and linear polarization along the lab $\hat{Z}$ axis: $\vec{E}(t) = \hat{Z} E_0 \exp\left(-\frac{t}{\sigma^2}\right) \sin(\omega_0 t)$, carrier frequency of $\omega_0 = 0.5 THz$ and a 1ps full width half max ($2(\ln 2)^{1/2} \sigma$ = 1ps). Thus $\vec{E}(t)$ is a perfectly anti-symmetric field with $\int_{-\infty}^{\infty} \vec{E}(t) dt = 0$ and with typical experimental parameters [10, 11, 12, 33]. The dipole interaction of the field and the molecule is described by the Hamiltonian: $\hat{H} = \hat{H}_0 - \hat{\vec{\mu}} \cdot \vec{E}$, where $\hat{H}_0 = \frac{\hat{L}_a^2}{2I_a} + \frac{\hat{L}_b^2}{2I_b} + \frac{\hat{L}_c^2}{2I_c}$ is the field-free asymmetric top rotational Hamiltonian, and $\hat{\vec{\mu}}$ is the dipole moment operator. The density matrix for the initial classical thermal ensemble is diagonal and consists of the initial rotational populations given by the Boltzmann distribution: $\frac{e^{-E_{JM\tau}/k_B T}}{Z}$ with temperature $T$, $Z = \sum_{J,M,\tau} e^{-E_{JM\tau}/k_B T}$ the partition function, and $E_{JM\tau}$ are the eigenenergies of the asymmetric top. $J$ and $M$ are the quantum numbers for the total angular momentum and its projection onto the lab Z-axis, respectively. For asymmetric rotors, the quantum number K, which corresponds to the projection of the angular momentum on the molecular fixed z-axis, is not a good quantum number and $\tau$ is introduced as the index that numerates the asymmetric top eignestates formed by a superposition of the symmetric top eigenstates following the textbook convention [34]. We start by calculating the number of rotational energy levels ($N_E$) as a function of temperature. Figure 1a presents the number of energy levels vs. the energy units of temperature. At $T = 40K$, the size of the density matrix that represents correctly the initial ensemble is $N_E$=500, thus an exact simulation using the full density matrix is computationally highly demanding already at this relatively low temperature. Existing methods for numerical propagation of the density matrix and its exact dynamics according to the Liouville-Von Neumann equation are inefficient as they scale as $N_E^3$ [35]. An alternative method that is commonly used [36] relies on propagating each and every state in the basis set of the asymmetric top $|JM\tau\rangle$ separately within the Schrödinger equation framework. The time dependent observables for each of these states are calculated throughout its dynamics, followed by their incoherent thermally weighted sum that yields the ensemble dynamics. Methods for wavefunction propagation such as the Chebychev or the Newton polynomial approximations scale as $N_E^2$ which makes them easier to handle [37]. However, as demonstrated in figure 1c for nonlinear molecules with three principal moments of inertia, $N_E$ scales as $T^{3/2}$ and the computation of the dynamics at temperatures above 100K with $N_E \sim 10^4$ becomes extremely challenging.

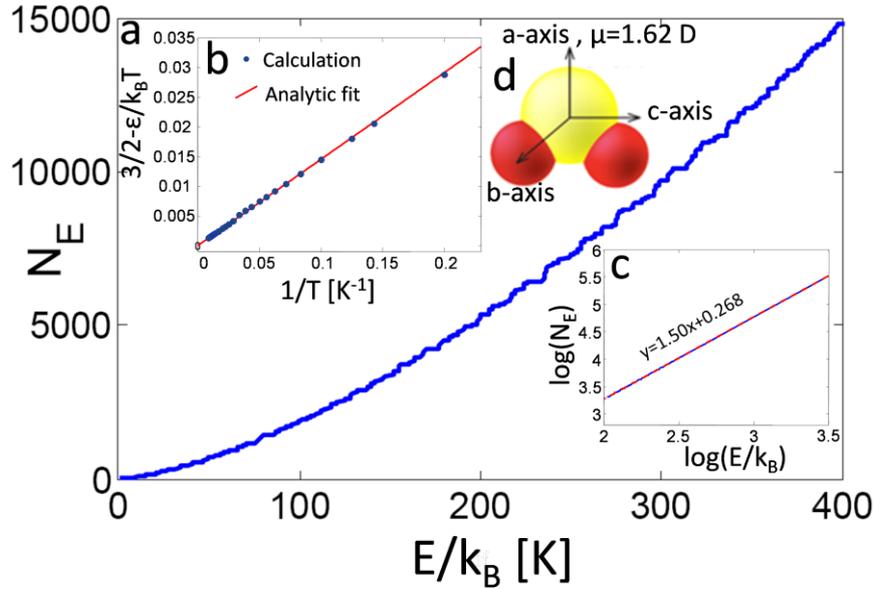

Figure 1: (color online) (a) Number of states as a function of the energy for the rotating $SO_2$ molecule. (b) Difference between the classical value of $3/2k_BT$ of the thermal ensemble and its actual value as a function of the inverse of the temperature. The exact and RPWF values are similar as shown by the solid red linear fit. (c) Same as (a) in logarithimic scale showing perfect scaling of $N_E \propto T^{3/2}$. (d) Sketch of the $SO_2$ molecule.

To overcome this computational barrier the Random Phase Wave Function (RPWF) method makes use of a limited ensemble of pure states that resembles a classical ensemble to enable an efficient representation of the mixed state [24,25]. A single RPWF is taken as:

$$\left| \vec{\theta}_k \right\rangle = \sum_{JM\tau} \sqrt{\frac{e^{-E_{JM\tau}/k_BT}}{Z}} e^{i\theta_{JM\tau}^k} \left| JM\tau \right\rangle \quad (1)$$

where the magnitude of the wavefunction amplitudes are given by the Boltzmann weighting and the phases $\theta_{JM\tau}^k$ are chosen randomly so that for large number of realizations k, the set of RPWF serve as a complete basis of states:

$$\lim_{k \to \infty} \sum_k \left| \vec{\theta}_k \right\rangle \left\langle \vec{\theta}_k \right| \cong \mathbf{1} \quad (2)$$

As in the exact approach, time dependent observables are computed for each of the RPWFs and an incoherent sum over many RPWF realizations which contain the right thermal weighting by construction (Eq. 1) yields the desired dynamics. While each RPWF $|\vec{\theta}_k\rangle$ is a coherent superposition of the asymmetric rotor eigenstates, *it is the randomization of the superposition phases that effectively cancels the overall initial coherence of the system*. The requirement for phase randomization makes the RPWF method more efficient as the size of the Hilbert space (and the number of random phases) increases, namely with the increase in temperature. This results in faster convergence, and as will be demonstrate in what follows, dramatically reduces the number of realizations *k* required at higher temperatures.

We now analyze the efficiency and ability of the RPWF method in simulating the THz-induced rotational responses of $SO_2$, by comparing to the exact dynamics calculation in which each and every eigenstate $|JM\tau\rangle$ is propagated separately within the Schrödinger equation framework, and the time dependent observables are calculated by their incoherent thermally weighted sum [36].

As a first step, we present the RPWF calculations of the *static* observables relevant to this study - the orientation and the alignment of the thermal isotropic ensemble of $SO_2$ molecules. Diagonal observables such as the energy, $\langle\psi|H|\psi\rangle$, are independent of the randomly chosen phase terms of the superposition's eigenstates since they do not couple between different eigenstates, therefore the energy calculated via the RPWF perfectly coincides with the exact calculation of the rotational energy, as is evident from figure 1b. Note that the rotational energy approaches the classical limit of $\frac{3}{2}k_BT$ with the increase of $T$. Moreover, the deviation of the calculated energy from the classical $\frac{3}{2}k_BT$ decreases linearly with $1/T$ [see ref. 38]. Fig. 2a compares the accuracy of the RPWF-obtained orientation $<\cos\theta>$ with the exact orientation. The angular distribution of a thermal ensemble is purely isotropic with an orientation value of $<\cos\theta>=0$. Being an off-diagonal observable, the orientation observable couples between different states of the rotational superposition, and thus depends on the phase factors, leading to an error in the RPWF calculated value. To characterize the error in the RPWF-calculated orientation, we calculate the absolute orientation squared $|\langle\psi|\cos\theta|\psi\rangle|^2$ for many RPWF realizations and plot in figure 2a the averaged inverse of the error $(\Delta\cos\theta)^{-1}\equiv 1/|\langle\psi|\cos\theta|\psi\rangle|^2$ as a function of the number of RPWF realizations $N_r$. This is repeated for different initial ensemble temperatures. In all of the calculations presented in this paper, the actual number of computed realizations was significantly larger than the noted $N_r$, in order to verify the convergence of the RPWF calculation for each given $N_r$. An exact orientation value of $\langle\cos\theta\rangle=0$ is obtained asymptotically $(\Delta\cos\theta^{-1}\to\infty)$ with increasing $N_r$ and the inverse square of the error scales linearly with $N_r$, thus the error scales as $\sim N_r^{-1/2}$ [24]. From Fig. 2a it is evident that as the temperature increases, the RPWF orientation converges faster to the exact (isotropic) value and thus fewer realizations are required. Figure 2c depicts $\alpha_1$, the slope of the inverse orientation error vs. the temperature on a logarithmic scale. From the linear fit we find that $\ln(\alpha_1) = \frac{3}{2}\ln(T)$, namely the error decreases with $T^{3/2}$ which corresponds to the number of initially populated states (see Fig. 1c). Figure 2b compares the molecular alignment $\langle\cos^2\theta\rangle$ obtained by the RPWF method to the exact value of the isotropic ensemble of $\langle\cos^2\theta\rangle = 1/3$. Also here, the error scales as $\sim N_r^{-1/2}$. The error in alignment decreases with $T^2$ as shown in Fig.2d, namely, faster than for the orientation. This is attributed to the fact that while orientation is a purely off-diagonal operator that depends on the phases of the adjacent $J$ states, the alignment operator consists of both off-diagonal and diagonal terms, namely, depends on both the randomly constructed coherences and the exactly constructed populations.

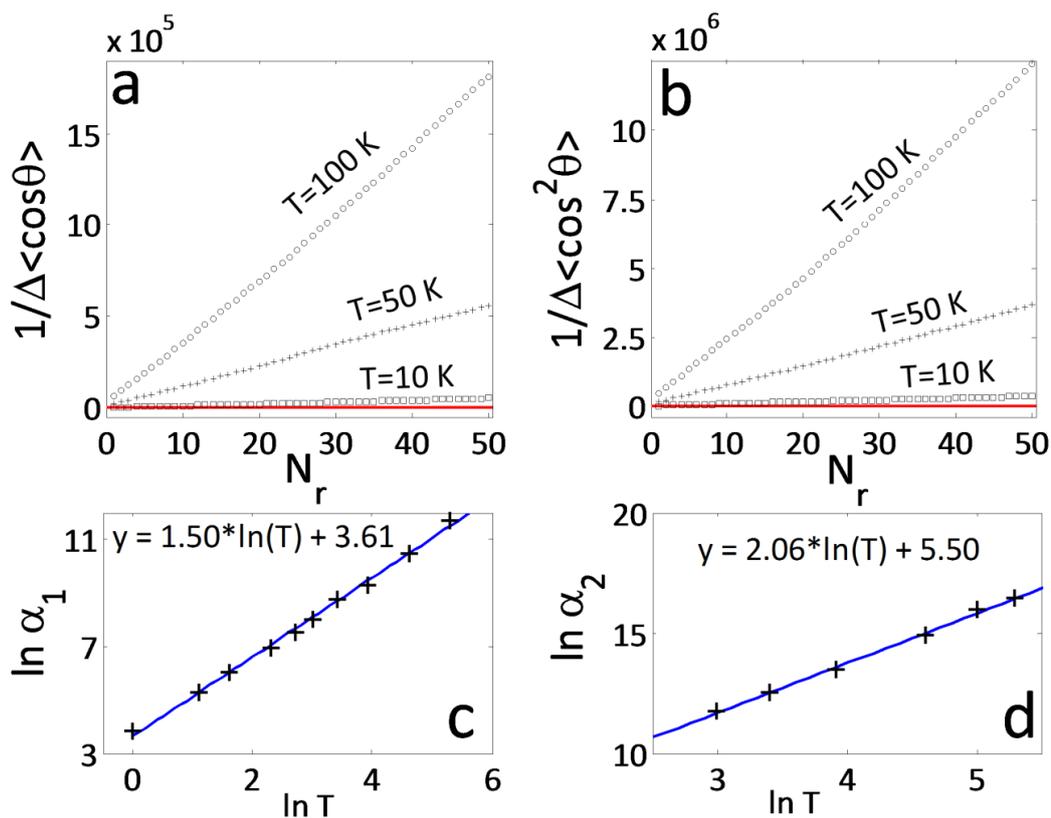

Figure 2: (color online) Static rotational observables calculated by the RPWF method for $SO_2$ ensembles at different initial temperatures. (a) and (b) shows the inverse error in the anisotropic orientation and alignment respectively for the isotropic thermal ensemble at 10, 50 and 100 Kelvin. (c) The slope of $\alpha_1$, the inverse error of the orientation, as a function of temperature on a logarithmic scale. (d) Same as panel c for the alignment ($\alpha_2$).

We proceed by exploring the accuracy and feasibility of the RPWF method in calculating the *field-induced orientation dynamics*. We set the peak intensity of the applied field to $I_0 = 2 \times 10^9 W/cm^2$, corresponding to a peak field of $1.2 MV/cm$, within the capability of current terahertz generation methods [see 39 and references therein]. In means of the effective order of excitation, such field strengths are in the weak to intermediate interaction limit for typical dipolar molecules.

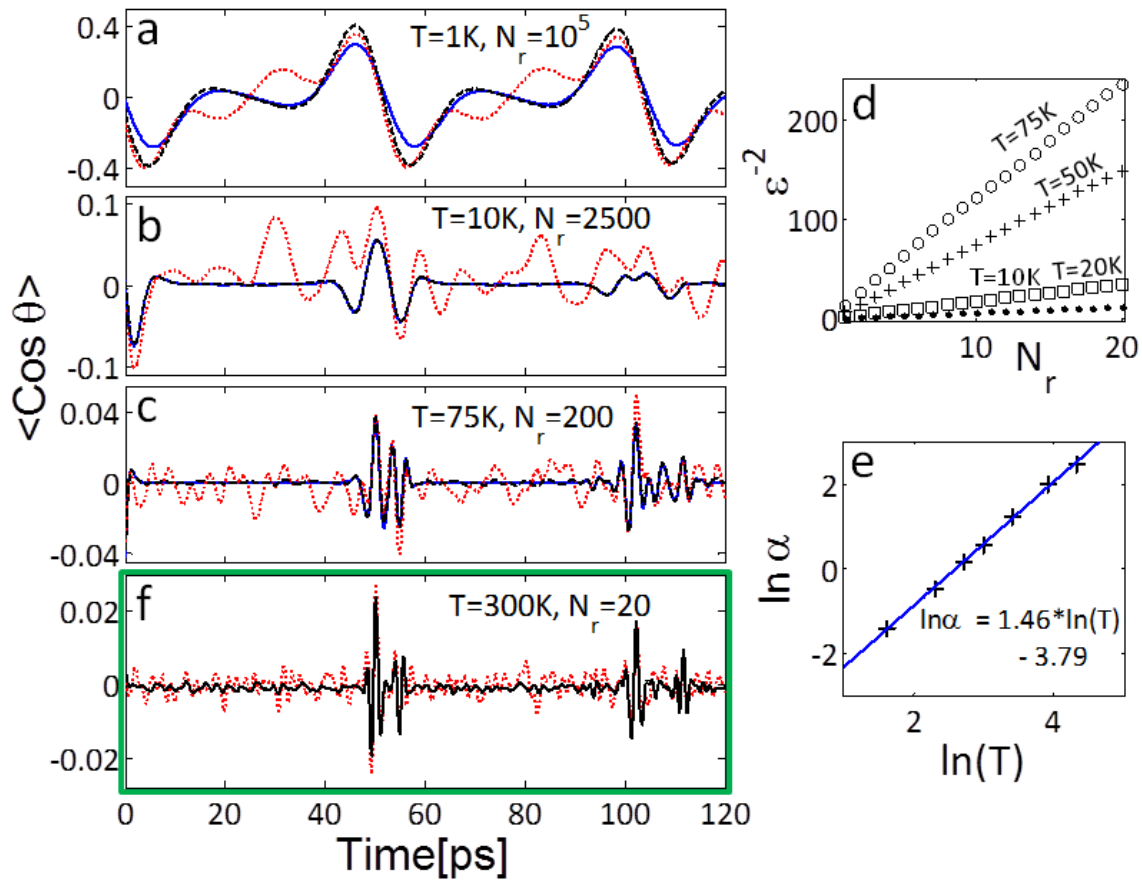

Figure 3 : (color online) Calculated orientation dynamics induced by a single-cycle THz pulse with peak field amplitude of 1.2 MV/cm. (a-c) : Molecular orientation dynamics induced by the THz pulse in thermal ensembles with initial temperature of 1, 10 and 75K respectively. The exact dynamics are depicted by solid blue lines and the RPWF dynamics are depicted by dashed black lines with $N_r$ noted in each figure. The red dotted lines show the orientation obtained by a single realization of the RPWF for comparison. (d) Averaged inverse of the error squared vs. $N_r$ for several initial temperatures of the ensemble. (e) Slopes of the linear fits as in Fig. (d) vs. temperature on a logarithmic scale. (f) RPWF calculation of the THz-induced orientation dynamics for $SO_2$ at 300K. Note that at ambient temperature, the exact calculation cannot be performed at reasonable computation time and thus the exact blue line is absent from the figure. At this high temperature and correspondingly large number of initial rotational states, the RPWF method reaches convergence already after 20 realizations, deduced here from the flatness of the orientation curve between fundamental revival times.

Figure 3a-c display the orientation of the molecules as a function of time subsequent to their interaction with a single-cycle THz field for the initial temperatures of T=1, 10 and 75K, respectively. The solid blue, dotted red and dashed black lines in each panel show the exact orientation, the orientation obtained with a single RPWF realization, and the orientation obtained with a large $N_r$, respectively. At T=1K (Fig 3a) the RPWF orientation modulations are synced with those of the exact calculation, but even after $10^5$ realizations, significant deviations of the RPWF peak amplitudes are observed. This results from the small number of initial states at low temperatures which corresponds to very limited phase randomization and decreased RPWF efficiency. For higher temperatures the method gains efficiency and a very good agreement with the exact signal is obtained for $N_r$ = 2500 at T = 10K (Fig 3b), and $N_r$=200 for T=75K (Fig. 3c). In order to quantify the error of the RPTW-obtained orientation dynamics we define an error function:

$$\varepsilon = \frac{1}{T_{rev}} \int_0^{T_{rev}} |S_{RP}(t) - S_{Ex}(t)|^2 \, dt \qquad (3)$$

where $S_{RP}$ and $S_{Ex}$ are the signals calculated using the RPWF and the exact method, respectively. We square the difference between the two obtained signals and average over the 120ps of the rotational evolution. The time span of the calculation was chosen to be larger than the longest period of SO$_2$ rotation given by $1/Cc$ where $C$ is the smallest rotational constant of SO$_2$ [in cm$^{-1}$] and $c$ is the speed of light. Fig 3d depicts the inverse error squared ($\varepsilon^{-2}$) vs. the number of realizations. Much like in the static observables, the convergence of the RPWF to the exact signal scales as $N_r^{-1/2}$, and the slope of $\varepsilon^{-2}$ scales with the temperature as the initial number of states, i.e., $\varepsilon^2 \sim T^{-3/2}$ as shown in Fig 3e. Taking both the initial temperature and the number of realizations into account, we deduce that the error scales as $\varepsilon \sim 1/\sqrt{N_r T^{3/2}}$, suggesting that for obtaining a converged signal for SO$_2$ at ambient temperatures, the number of realizations required is $N_r$~20. This is shown in Fig. 3f where the orientation signal at T = 300K was calculated by averaging only 20 RPWF realizations. We use the flatness of the signal at times between the rotational wavepacket revivals as an indication for convergence since an exact calculation of such a large system is not feasible under any exact method that is currently used.

Driven by the improved RPWF convergence at high initial temperatures as described above, we continue by examining the RPWF performance as a function of the number of the eigenstates involved in the RPWF. Specifically, we start with the same initial ensemble temperatures as in the previous section and increase the number of participating eigenstates with the help of higher field strengths that induce population transfer to higher rotational energy states. For this task we've calculated the orientation dynamics induced by the THz pulse with a peak intensity $I_0 = 1 \times 10^{11} W/cm^2$ that is 50 fold stronger than in the previous section. Such high field strengths ($\sim 10 MV/cm$)[40] induce multiple dipole transitions and significantly extend the number of populated states. Nevertheless, the number of randomly chosen phases for each given initial temperature remained the same as in the previous sections. To ensure a clean comparison between the convergence rates of the RPWF for weak and strong field interactions, we restrict the calculation to the dipole interaction term, as done for the weak field case. In general, at high field strengths and depending on the ratio between the anisotropic polarizabillity and the dipole moment, $E_0 \Delta\alpha / \mu$, nonlinear interaction terms may become significant and should be included in the interaction Hamiltonian. For the field and molecular parameters used here, $E_0 \Delta\alpha / \mu \sim 0.01$, and the dipole interaction dominates the dynamics. Figures 4a-d depict the orientation dynamics induced by the strong field for T = 3, 10, 75 and 300K. By comparing to the weak field orientation dynamics of Figs. 3a-c, one finds that for the high temperature case (T=75K), the strong field orientation (Fig. 4c) converges faster than the weak field (Fig 3c), with the $\varepsilon^{-2} \sim 200$ values reached with $N_r \sim 12$ and $N_r \sim 17$ respectively. Even the single RPWF realization depicted by the dotted red line in Fig 4c coincides almost perfectly with the exact modulations (solid blue curve) of the orientation at the fundamental revival times. Another demonstration for the improved strong-field convergence of the method is evident in Fig. 4d by the effectively flat orientation baseline obtained with only 10 RPWF realizations, compared to ~20 realizations required for the weak field case (Fig. 3d).

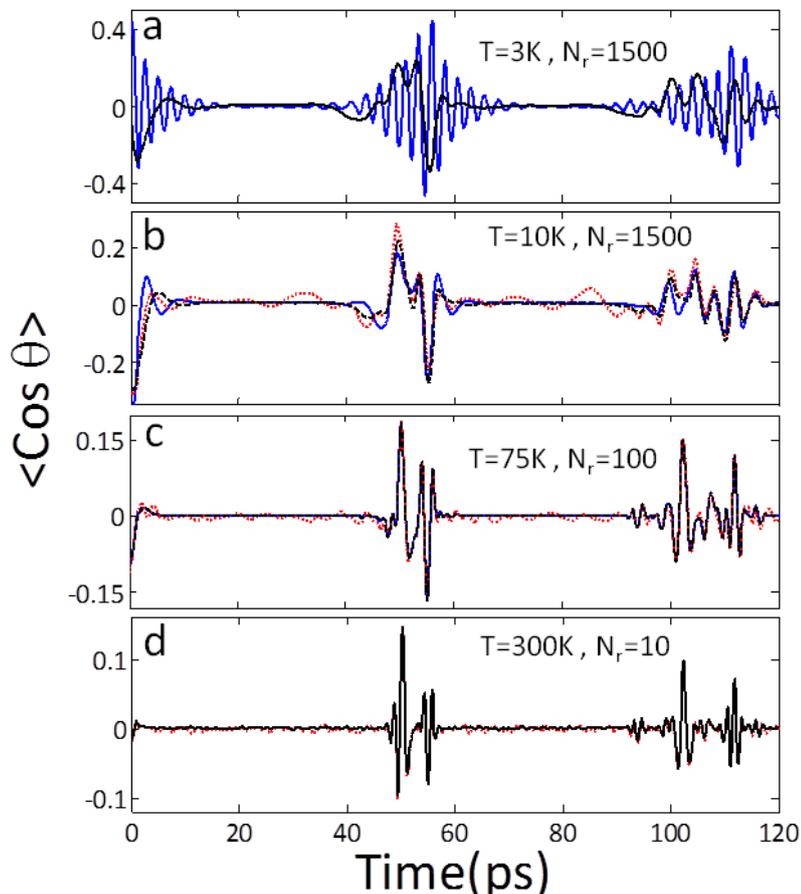

Figure 4: (color online) (a-d) Orientation dynamics induced by a strong THz field with peak intensity of 1X10$^{11}$ W/cm$^2$ in SO$_2$ ensembles at 3K, 10K, 75K and 300K respectively. The exact, RPWF (with $N_r$ given in the figure) and RPWF with a single realization are shown by the solid blue, dashed black and dotted red curves respectively.

However, while the RPWF gains efficiency and converges faster at higher temperatures and stronger field strengths as shown in Fig 4 b-d, it totally fails to predict the strong-field orientation induced at the low temperature case of Fig. 4a, regardless of the number of realizations. In this regime, the exact method dynamics exhibit a strong, multicycle modulation which results from multiple field-dipole interactions via which the rotational populations are transferred to angular momentum levels (J quantum numbers) much higher than the initial thermal ensemble through ladder climbing. In this extreme case, the initial number of purely random phases that were chosen arbitrarily is very small, and most of the phases of the wavefunction constituents following the strong interaction had in fact evolved from the few initial states, thus they are not purely random. This limited randomization of phases within each single realization leads to failure of the RPWF in predicting the correct orientation dynamics as observed in Fig. 4a. We note that for such low ensemble temperatures, the exact dynamics calculations are favorable and the RPWF method does not provide any significant advantage whatsoever. Detailed examination of the strong field induced modulation of Fig. 4a is under way.

In conclusion, we've utilized the RPWF method for computing the field-induced rotational dynamics of an asymmetric top molecular ensemble at different temperatures and field strengths, well beyond the capabilities of existing computation methods. To the best of our knowledge, the RPWF method is compared here, for the first time, to exact numerical dynamics and asymptotic scaling laws with the temperature are derived. The efficiency of the RPWF method was found to *increase* with the ensemble's temperature and number of initial states. At low temperatures, the convergence of the RPWF to the exact dynamics requires a large number of random realizations. However, at high ensemble temperatures and corresponding large number of initial states, convergence to the exact dynamics is rapidly reached with only a few realizations required, thus enabling accurate simulations with standard computational resources.

Our results for $SO_2$ show that the computational cost of the exact and RPWF methods becomes comparable at $T \sim 30K$. At ambient temperatures and above, the RPWF method is, to the best of our knowledge, the only option to compute the rotational dynamics in a reasonable effort. Moreover, for high enough temperatures, the computational cost for the method becomes temperature independent, as both the increase of accuracy and the number of initial states scale as

$T^{3/2}$. Thus, 20 RPWF realizations are sufficient for retrieving accurate rotational dynamics of the SO$_2$ at room temperature and at $T \sim 10^3 K$ a single realization would suffice. Furthermore, the RPWF was found to converge faster with the increase of the field intensity, as long as the dynamics is not dominated by a large number of successive coherent THz field-dipole interactions.

The exact same RPWF approach can be used to calculate non-resonant optical-induced alignment in thermal molecular ensembles and can be extended to include multiple pulses / shaped laser fields. In addition, the RPWF is advantageous in calculating incoherent processes such as dephasing due to collisions and field instabilities as it inherently provides multiple realizations which are immanent in the method and hence the Stochastic Schrödinger Equation can be implemented.

On a more general level, the broader field of molecular dynamics and associated coherent rotational control aims at exploring large molecular species of chemical and biological importance. Such large asymmetric molecules have high density of rotational states and correspondingly consist of hundreds and possibly thousands of initially populated states even at molecular beam temperatures. Namely, for larger molecules the RPWF method is expected to be far more efficient than existing molecular dynamics methods, and probably the only option for predicting the field-induced (THz and/or near-IR) rotational dynamics.

We wish to acknowledge Ronnie Kosloff (HUJI) for stimulating discussions. This work was supported by the Binational Science Foundation grant no. 2012021 (S. K), by the Israeli Science Foundation grant no.1065/14 (S. F.) and in part by the Marie Curie CIG grant no. 631628 (S. F.).

† Email: shimshonkallush@braude.ac.il
‡ Email: sharlyf@post.tau.ac.il